\documentclass[conference]{IEEEtran}
\IEEEoverridecommandlockouts
\usepackage{amsmath}
\usepackage{color,soul}
\usepackage{multicol}
\usepackage{balance}
\usepackage{array}
\usepackage{verbatim}
\usepackage{stfloats}
\usepackage{amsthm} 
\usepackage{multirow}
\usepackage{amssymb}
\usepackage{float}
\usepackage{nccmath}
\usepackage{graphicx}
\usepackage{setspace}
\usepackage[algo2e, linesnumbered, ruled, vlined]{algorithm2e}
\usepackage{mathtools}
\usepackage{booktabs}

\usepackage{mwe}
\usepackage{cite}
\usepackage{amsmath,amssymb,amsfonts}
\usepackage{graphicx}
\usepackage{textcomp}
\usepackage{xcolor}
\usepackage{physics}
\usepackage{bm}
\usepackage[utf8]{inputenc} 
\usepackage{kotex} 
\usepackage[linesnumbered,ruled]{algorithm2e}
\usepackage{cite}
\usepackage{amsmath,amssymb,amsfonts}
\usepackage{algorithmic}
\usepackage{graphicx}
\usepackage{textcomp}
\usepackage{xcolor}

\newcommand{\BfPara}[1]{{\noindent\bf#1.}\xspace}

\usepackage[normalem]{ulem}

\usepackage{hyperref}
\usepackage{kotex}
\usepackage{tikz}

\newcommand{\tred}[1]{{\color{red}{#1}}}

\def\BibTeX{{\rm B\kern-.05em{\sc i\kern-.025em b}\kern-.08em
    T\kern-.1667em\lower.7ex\hbox{E}\kern-.125emX}}
    
\begin{document}

\title{3D Scalable Quantum Convolutional Neural Networks for Point Cloud Data Processing in Classification Applications}

\author{
\IEEEauthorblockN{Hankyul Baek, Won Joon Yun, and Joongheon Kim}
\IEEEauthorblockA{\textit{Department of Electrical and Computer Engineering, Korea University, Seoul 02841, Republic of Korea} \\
\texttt{67back@korea.ac.kr}, \texttt{ywjoon95@korea.ac.kr}, \texttt{joongheon@korea.ac.kr}}
}

\maketitle

\begin{abstract}
With the beginning of the noisy intermediate-scale quantum (NISQ) era, a quantum neural network (QNN) has recently emerged as a solution for several specific problems that classical neural networks cannot solve. Moreover, a quantum convolutional neural network (QCNN) is the quantum-version of CNN because it can process high-dimensional vector inputs in contrast to QNN. However, due to the nature of quantum computing, it is difficult to scale up the QCNN to extract a sufficient number of features due to barren plateaus. Motivated by this, a novel 3D scalable QCNN (sQCNN-3D) is proposed for point cloud data processing in classification applications. Furthermore, reverse fidelity training (RF-Train) is additionally considered on top of sQCNN-3D for diversifying features with a limited number of qubits using the fidelity of quantum computing. Our data-intensive performance evaluation verifies that the proposed algorithm achieves desired performance.
\end{abstract}

\begin{IEEEkeywords}
Point Cloud Classification, Quantum Convolutional Neural Networks, Quantum Machine Learning.
\end{IEEEkeywords}

\section{Introduction}\label{sec: Inbtroduction}

A point cloud is expected to be desirable three-dimensional sensory data and it is widely and actively used in various fields, including robotics~\cite{pomerleau2015review} and autonomous vehicles~\cite{cui2021deep,t-its1,JCN1}. In addition, the point clouds are alternative data to precisely analyze and rebuild our surrounding real-world situations~\cite{yue2018lidar}. 
Currently, millions of point cloud vertices per second can be produced by laser imaging detection and ranging (LiDAR) sensors. Furthermore, according to \textit{Global Market Insights} \cite{GMI}, the market size of \textit{point cloud processing related media services} will reach more than 15 billion dollars in 2030. 

Spurred by the advance of LiDAR, recent studies on point cloud processing have shown that neural network (NN) can be an alternative solution for various tasks such as classification~\cite{hackel2017semantic3d}, object detection~\cite{zhou2018voxelnet,guo2020deep}, and segmentation~\cite{landrieu2018large}. 
However, due to abundant spatial geometric information and the massive size of the point cloud, operating the point cloud processing is still challenging~\cite{qin2019pointdan}. A point cloud usually has around 100k vertices, and each vertex consists of a lot of information, \textit{i.e.}, color and Cartesian coordinates. Therefore, classical computing methodologies are obviously harsh to utilize the point cloud in real-time and even jammed when they utilize NN. 

Quantum computing can be a reasonable solution to these challenges. The quantum computing utilizes a quantum bit (qubit), the counterpart in quantum computing to the binary bit of classical computing. Due to the characteristic of qubit that ranges from $0$ to $1$, not exact integer $0$ or $1$ in classical computing, quantum computing can enlarge its computation capability on an exponential scale~\cite{DBLP:conf/icde/BaiJCRWYH22,aimlab2022arvix,pieee202105park,kong2020review}. 
In addition, parallelization is one of the beauty of quantum computing. Thus, quantum algorithms have shown that they are able to solve several NP-hard problems in polynomial time, such as Shor algorithm~\cite{shor1994algorithms} and Grover search algorithm~\cite{grover1996fast}, those are physically impossible in classical approaches~\cite{aimlab2022icdcs}.
Therefore, quantum computing can outperform classical algorithms in terms of processing speed for some problems under specific conditions~\cite{spl01,spl02}.

In this paper, we focus on grafting quantum machine learning (QML) methodologies to point cloud data processing which requires tremendous computational costs~\cite{liu2019pvcnn}. 
In na\"{i}ve and straightforward approach to process point cloud data, increasing the number of qubits or gates can be intuitively considered. However, the trainability issue occurs because the number of local minima (i.e., barren plateau) is proportion to the exponential number of gates in QNN~\cite{mcclean2018barren}. Therefore, this approach is not considerable. 
Fortunately, it has been proven that the barren plateaus can be eliminated in quantum convolutional neural network (QCNN)~\cite{pesah2021absence}. Furthermore, image processing with QCNN shows better feasibility when QCNN is used together with classical NN, i.e., fully connected network (FCN)~\cite{Oh2020QCNN}, which can be called a quantum-classical hybrid classifier. It has been experimentally proven that hybrid QML achieves considerable performance in classification tasks~\cite{PhysRevLett.122.040504}. Thus, it is better to use hybrid QCNNs in point cloud processing applications.

On top of these current research progresses, there still remain important questions on QCNN methodologies, i.e., \textit{(i) how to upload tremendous classical data into QCNN?}, \textit{(ii) how to train QCNN efficiently?}, and \textit{(iii) how to make the complexity of QCNN be proportioned to performance?}
In order to answer these questions, we aim to extend a 3D voxelized version of scalable QCNN under the consideration of barren plateaus, data uploading issues, and efficient training methods. 
First of all, we propose a 3D voxelized version of scalable QCNN, i.e., 3D scalable QCNN (sQCNN-3D). 
Moreover, we alleviate the data uploading issue by adopting data-reuploading \cite{P_rez_Salinas_2020}. Furthermore, a new scaling strategy is proposed for leveraging the various sizes of quantum convolutional filters. Finally, in order to realize efficient training, we propose a 3-dimensional reverse fidelity-train (3D RF-Train), which let sQCNN-3D fully utilize the point cloud's intrinsic features while utilizing only a limited number of qubits. Our proposed methods can not only resolve the aforementioned challenges but also answer the questions mentioned above.

\BfPara{Contributions} The major contributions of the research results in this paper can be categorized as follows.
\begin{itemize}
    \item First of all, a novel scalable QCNN architecture for extensible 3D data processing, i.e., sQCNN-3D, in order to achieve scalability while pursuing quantum supremacy and also avoiding barren plateaus.
    \item Moreover, an additional sQCNN-specific training algorithm, i.e., \textit{RF-Train}, in order to extract the intrinsic features with a finite number of qubits.
    \item Furthermore, a new scaling strategy is also proposed which unleashes the potential of sQCNN-3D. More filters are corroborated in order to induce higher accuracy.
    \item Lastly, data-intensive experiments are conducted to corroborate the superiority of sQCNN with RF-Train, in ModelNet and ShapeNet, widely used in the literature.  
\end{itemize}

\BfPara{Organization} The rest of this paper is organized as follows. Sec.~\ref{sec:preliminaries} briefly introduces the differences between CNN and QCNN; and the concept of quanvolution. Sec.~\ref{sec:point cloud processing} presents our proposed 3D scalable QCNN for point cloud processing in classification. Sec.~\ref{sec:experiment} evaluates the performance of the proposed algorithm. Lastly, Sec.~\ref{sec:conclusions} concludes this paper.

\section{Preliminaries}\label{sec:preliminaries}

\subsection{Point Cloud Processing with Classical CNN} 
There are two main categories in point cloud processing, i.e., point-wise processing~\cite{qi2017pointnet} and voxel-based processing~\cite{wu20153d,maturana2015voxnet}, and the later one is the major trend in the literature.
A classical CNN for the voxel-based processing is mainly composed of four procedures, i.e., embedding, convolution, pooling, and prediction.

\BfPara{Embedding}
In contrast to a 2D image, each point cloud in 3D data has a huge number of vertices. Thus, a set of vertices in each point cloud is embedded as features to reduce the computational complexity as well as improve the robustness for perturbation. PointGrid~\cite{le2018pointgrid} embeds each set of vertices into voxel grids to achieve sophisticated local geometric features; and DGCNN~\cite{wang2019dynamic} extracts edge features from a set of vertices to incorporate local neighborhood information.

\BfPara{Convolution} After embedding the high dimensional point cloud data into features, the features can be convoluted by a set of filters. With trainable parameters, filters slid across each axis of features, i.e., named \textit{input features}. The dot products between the input features and filters are computed at every spatial position.

\BfPara{Pooling} This pooling computation is conducted to reduce the dimension of the convolved input features. This is usually considered as a critical part of point cloud CNN-based models because it speeds up the subsequent convolution layer computation~\cite{DBLP:conf/cvpr/WangYMHHX16}. In addition, it allows these models to learn representations invariant to minor translations.

\BfPara{Prediction}
For the prediction, the fully connected layers receive feature information which is derived from convolution layers and pooling layers. According to the universal approximation theorem~\cite{DBLP:journals/corr/abs-2012-13882}, fully connected layers are allowed to make a prediction. By optimizing the objective function (e.g., cross-entropy loss or mean-squared error loss), conventional classical CNN can achieve the prediction to the desired classes.

\subsection{Quantum CNN}\label{sec:quanvolution}
QCNN is the quantum version of CNN, which leverages a parameterized quantum circuit (PQC) as convolutional filters. With the QCNN  which utilizes \textit{quanvolutional} filters~\cite{DBLP:journals/qmi/HendersonSPC20}, spatial information can be exploited with particular characteristics. Note the definition of quanvolution is explained later.

\BfPara{Basic Quantum Operation}
All quantum-related notations and their operations are represented as Dirac-notation. In contrast to classical computing, a qubit can have two possible states denoted as $|0\rangle$ and $|1\rangle$~\cite{guan2021quantum}. 
The quantum state in an $q$-qubit system is defined as $2^{q}$ possible bases and their probability amplitudes, i.e., $\ket{\psi} \triangleq \sum^{2^{q}}_{k=1} \alpha_k \ket{k}$ where $\ket{k}$ denotes a basis in Hilbert space, $\forall q \in \mathbb{N}[1,\infty)$ and $\sum^{2^{q}}_{k=1} |\alpha_k|^2 = 1$. The operation of quantum state $\ket{\psi}$ is represented with unitary matrix $U$, and its operation is expressed as $\ket{\psi} \leftarrow U \cdot \ket{\psi}$. 
Note that the quantum state is not deterministic. 
Therefore, the quantum state is transformed into classical data only with its measurement; and the measurement of the quantum state is represented as a set of projection matrices $\mathcal{M} \triangleq \{\mathbf{M}_k\}^{q}_{k=1}$. 
This paper uses the measurement matrix $\mathbf{M}_k=\mathbf{I}^{\otimes k-1}\otimes \mathbf{Z} \otimes \mathbf{I}^{\otimes q -k}$, $\forall k \in [1, q]$, where $\mathbf{I}$ denotes the identity matrix and $\mathbf{Z} = \begin{bmatrix}
1 &  0 \\
0 & -1
\end{bmatrix}$. Then, the classical output (called observable) is obtained as follows, 
$\langle O_k \rangle =\langle \psi | \mathbf{M}_k |\psi \rangle$. 
Here, the measurement is an activation function. 
In other words, the unitary matrix and measurement operation are mapped to linear operation and activation function, respectively, which leads to QML feasible~\cite{hamamura2020efficient}.

\begin{figure}[t!]
\centering
\includegraphics[width=\linewidth]{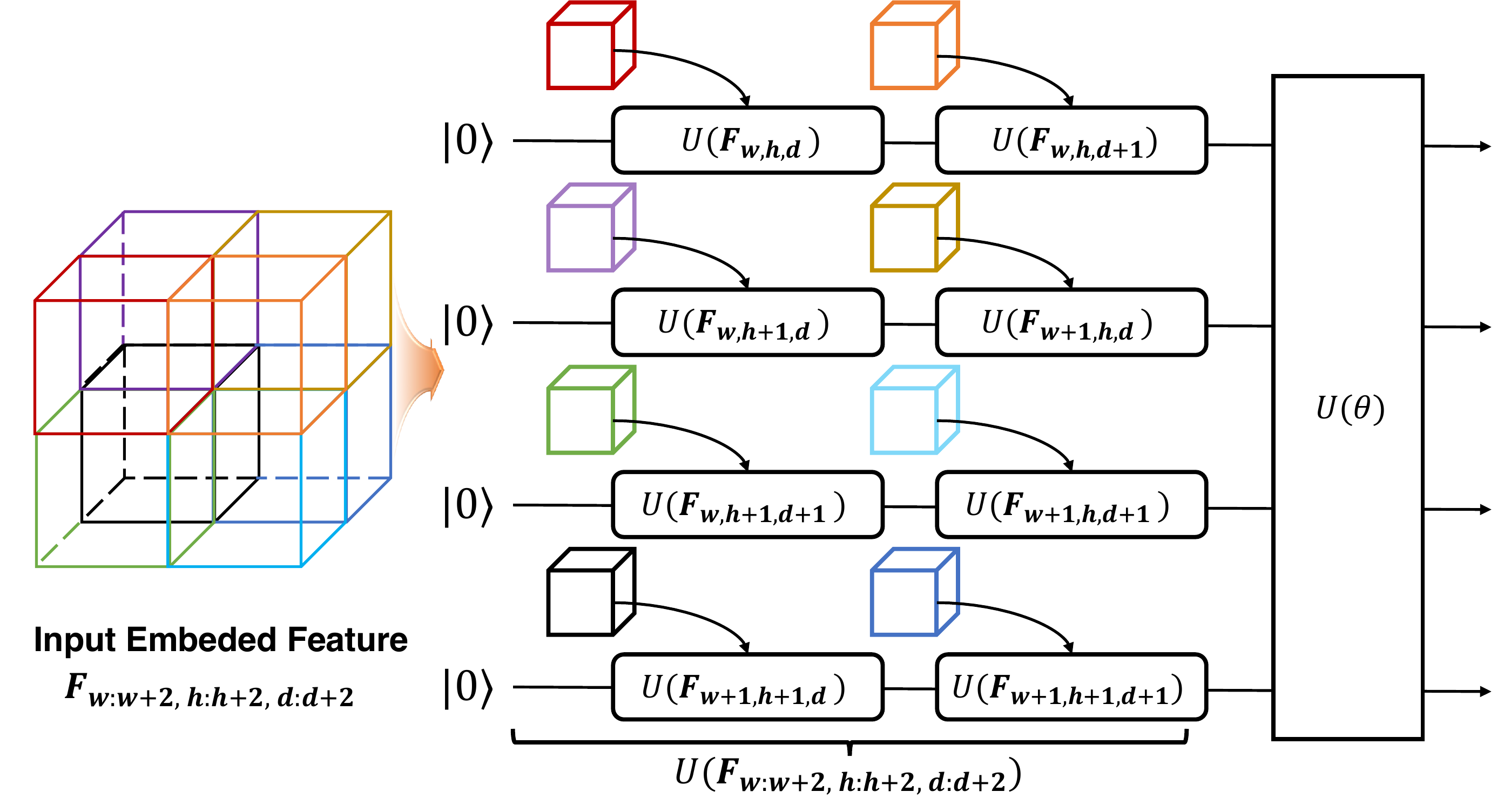}
\caption{An illustration of multi-qubit reuploading in sQCNN-3D ($q$=4, $\kappa$=2).}\label{fig:data_reuploading}
\end{figure}

\BfPara{Quanvolution} The quanvolutional filter is defined to follow the two consecutive procedures, i.e., (i) data encoding-processing and (ii) measurement. 

\begin{enumerate}
    \item \textit{Data Encoding-Processing:} To use quanvolutional filter with classical data, the encoding strategy should be considered where classical data is transformed into quantum states.
In QML research, several encoding strategies are discussed such as basis-encoding, amplitude-encoding, and angle-encoding~\cite{PhysRevLett.122.040504}. Among them, it is experimentally known that the angle-encoding presents the best performance in QML applications. Especially, the angle-encoding with data reuploading shows the high performance when the classical data $\mathbf{x}$ are encoded to quantum states, which makes the quantum states efficiently utilize the Hilbert space using trainable parameters $\bm \theta$~\cite{P_rez_Salinas_2020}.
This process can be expressed as follows,
\begin{equation}
    |\psi\rangle = \prod_{l=0}^{ \lceil \text{size}(\mathbf{x}) / q\rceil }U(\theta_m)U(\mathbf{x}_{q \cdot l:q\cdot(l+1)})|0\rangle^{\otimes q}\label{eq:data_reuploading}, 
\end{equation} 
where $U(\cdot)$ and $\text{size}(\cdot)$ stand for the unitary operation and the vector size of input, respectively.
Because the number of variables in input data is larger than the number of qubits $q$ in a modern noisy intermediate-scale quantum (NISQ) era, the input data $\mathbf{x}$ are splitted into $\mathbf{x}_{q\cdot l:q\cdot(l+1)}$ with the interval of $q$, $\forall l \in \left [1, \lceil \text{size} (\mathbf{x}) / q \rceil\right]$. Note that $\mathbf{x}_{q\cdot l:q\cdot(l+1)}$ and $\theta_l$ denote the input vector which is composed by the first $q \cdot l $ to $q \cdot (l+1)$ elements and the trainable parameters where $\forall \theta_l \in \bm \theta$. 
Similarly, the encoded quantum states are processed with PQC.
    \item \textit{Measurement:} The measurement operation has the role of not only pooling but a representation of spatial feature information, which is composed of various channels~\cite{DBLP:journals/qmi/HendersonSPC20}. 
In this paper, we encode classical data to quantum states by using this angle encoding with a data reuploading strategy. In addition, we regard the measurement operation as pooling and spatial representation.
\end{enumerate}

\section{3D Scalable Quantum Convolutional Neural Networks (sQCNN-3D) for Point Cloud Processing}\label{sec:point cloud processing}

This section presents the motivation, architecture, training algorithm (i.e., RF-Train), and overall algorithmic procedure of our proposed sQCNN-3D, respectively.

\subsection{Motivation}\label{sec:Motivation}

This paper designs sQCNN-3D for point cloud data processing in classification applications, based on following motivations. First of all, classifying point cloud data requires high computational complexity because typical point clouds in many applications contain approximately 100k points and each vertex in each point cloud uses 15 bytes~\cite{zhou2018voxelnet}. Second, various variational quantum algorithms suffer from vanishing gradients due to many qubits utilization. As discussed in~\cite{mcclean2018barren}, the variance of the derivative vanishes exponentially as the number of qubits increases. Therefore, it is obvious that leveraging a quantum circuit while using only a few qubits is essential during training point cloud classifiers.

\begin{figure}[t!]
\centering
\includegraphics[width=\columnwidth]{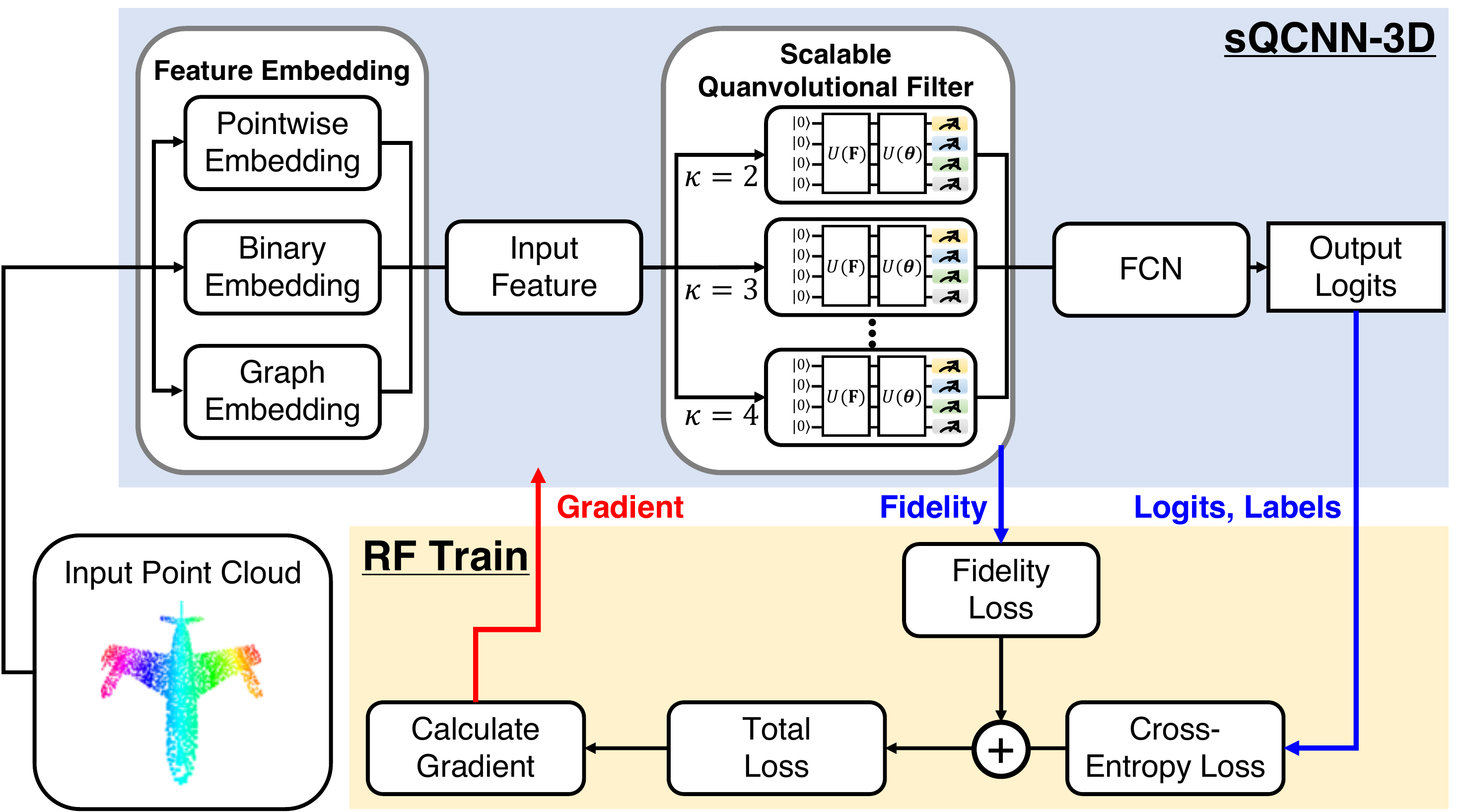}
\caption{The pipeline for processing point clouds with sQCNN-3D.} \label{fig:Model ARchi}
\end{figure}

\subsection{sQCNN-3D Architecture}\label{sec:archi}
The architecture of sQCNN-3D is as shown in Fig.~\ref{fig:Model ARchi} (blue box) where it consists of two major components as follows.

\BfPara{Feature Embedding} We consider a point cloud that spans in a 3D Euclidean space, which is defined as $\mathcal{P}\triangleq \{(x_i,y_i,z_i)|i=1,\cdots,|\mathcal{P}|\}$, and its $i$-th element (i.e., vertex) is denoted as $p_i$. We subdivide the 3D space into voxels $\mathcal{V}\triangleq\{(\hat{x}_{j}, \hat{y}_{j}, \hat{z}_{j})|j=1,\cdots,|\mathcal{V}|\}$, where its $j$-th voxel is denoted as $v_j$.
Because this paper only considers the geometric features of point cloud $\mathcal{P}$ in voxels $\mathcal{V}$, we obtain a geometric feature set which is denoted as $\mathcal{F}= g (\mathcal{P} \odot \mathcal{V}) \equiv \{f(v_j)| j = 1,\cdots, |\mathcal{V}|, \forall f(v_{j}) \in \{0,1\}\}$.
Here, the operator $\odot$ denotes that each $p_i$ is in the area of voxel $v_j$. The geometric feature extraction function $g(\cdot)$ is defined according to the target geometric feature \cite{hinks2013point}, i.e., Euclidean distance, structural similarity, and binary encoding~\cite{hinks2013point}.
We set the range of $\mathcal{P}$ and $\mathcal{V}$ as $(W, H, D) \in \mathbb{R}^3$ and ($\hat{W}, \hat{H}, \hat{D}) \in \mathbb{N}^3$. 
From the binary encoding in~\cite{hinks2013point}, $f(v_{j})=1$ if $n_j \geq T$ (otherwise, $f(v_{j})=0$) where $n_j$ is the number of vertices that satisfies 
$\sqrt{|x_i \cdot \frac{\hat{W}}{W}  - \hat{x}_j|^2} \leq 0.5$, 
$\sqrt{|y_i \cdot \frac{\hat{H}}{H}- \hat{y}_j|^2} \leq 0.5$ and 
$\sqrt{|z_i \cdot \frac{\hat{D}}{D}- \hat{z}_j|^2} \leq 0.5$, $\forall i \in \mathbb{N}[1,|\mathcal{P}|]$, $\forall j \in \mathbb{N}[1, |\mathcal{V}|]$. Note, $T$ is a threshold to determine whether $v_j$ is active or not.

\BfPara{3D Data Reuploading} Inspired by~\cite{P_rez_Salinas_2020}, we aim to embed classical bits into quantum states using only a limited number of qubits. In contrast to classical QCNN that embeds each data into a qubit respectively, sQCNN-3D solves the huge dimensionality problem in point cloud operation through data reuploading. Fig.~\ref{fig:data_reuploading} shows the data reuploading process of sQCNN-3D. We define the input embedded feature $\mathbf{F}$ which has the size of $ c_{in} \times \hat{W} \times \hat{H} \times \hat{D}$. 
We interchangely use the notation $\mathbf{F}_{w:w+\kappa, h:h+\kappa, d:d+\kappa}$ to represent the patch of which origin and the kernel size are $(w,h,d)$ and $\kappa$, respectively. Note that $\mathbf{F}_{w:w+\kappa, h:h+\kappa, d:d+\kappa}$ is a vector with the size of $\kappa^3$.
We encode the patch $\mathbf{F}_{w:w+\kappa, h:h+\kappa, d:d+\kappa}$ with \eqref{eq:data_reuploading} where $\forall w \in \mathbb{N}[1, \hat{W}], \forall h \in \mathbb{N}[1, \hat{H}], d \in \mathbb{N}[1, \hat{D}]$.

\begin{figure}[t!]
\centering
\includegraphics[width=.85\columnwidth]{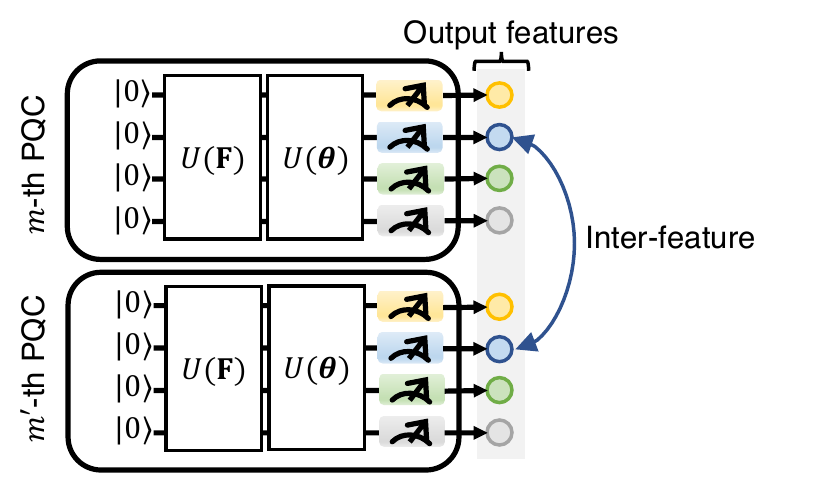}
\caption{The meaning of terminology in RF-Train.}
\label{fig:inter}
\end{figure}

\BfPara{Scalable 3D Quanvolutional Filter} Inspired by~\cite{baek2022scalable}, we aim to design the 3D version of scalable quanvolutional filters. Fig.~\ref{fig:Model ARchi} illustrates the architecture of 3D scalable filters. In contrast to QCNN, which has limitations in extracting numerous features due to barren plateaus, sQCNN-3D scales up the filter by increasing the number of filters \textit{(note that it does not increasing the number of qubits)}. By feed-forwarding each PQC architecture, the 3D quanvolutional filter transforms $\mathbf{F}_{w:w+\kappa, h:h+\kappa, d:d+\kappa}$ to an observable $\langle O_{w,h,d,c} \rangle$ from a 3D matrix of scalars, where $c$ denotes the output channel index, $\forall c \in \mathbb{N}[1,c_{out}]$. 
In order to extract various feature channels from the input patch, our proposed quanvolutional filter mainly consists of multiple layers of controlled-U (CU) gates. The operations of the CU gate in two qubits can be expressed as 
$\begin{bmatrix}
    I & 0\\
    0 & U
    \end{bmatrix}$, 
where $U$ is illustrated as 
$U=\begin{bmatrix}
    u_{00} & u_{01}\\
    u_{10} & u_{11}
\end{bmatrix}$. 
After processing with quanvolutional filter, akin to classical CNN, sQCNN-3D achieves a feature corresponding to the measurement value of each qubit in filters. However, the scalability is not straight-forwarded yet. Thus, the quanvolution filters are compared with the various filter-sizes $\kappa$. This strategy enables sQCNN-3D to achieve various spatial features, which results in performance improvements.

\subsection{RF-Train}\label{sec:rftrain} 
In contrast to classical QCNN, sQCNN-3D can increase the number of extracted features while maintaining the number of qubits in a PQC. To achieve the full advantage of sQCNN-3D, we use RF-Train~\cite{baek2022scalable}. 
In this paper, we consider $M$ PQCs and each PQC is composed of $q$ quanvolutional filters. 
Suppose that two PQCs exist, i.e., $m$-th and $m'$-th PQCs whose density operators are denoted as $\rho =|\psi_m\rangle \langle\psi_{m}|$ and $\sigma =|\psi_{m'}\rangle \langle\psi_{m'}|$, where $\forall m, m' \in \mathbb{N}[1,M]$. 
Denote $\langle O_k \rangle_{m}$ and $\langle O_k \rangle_{m'}$ as the $k$-th qubit's observables of $m$-th and $m'$-th PQC, respectively, where $\forall k \in [1, q]$.
In quantum information theory~\cite{bennett1998quantum}, the similarity between $|\psi_{m}\rangle$ and $|\psi_{m'}\rangle$ can be measured in terms of fidelity $\Omega (\psi_{m}, \psi_{m'}) = {(\Tr(\sqrt{{\sqrt{\rho}\sigma\sqrt{\rho}}}))}^2$, where $\Tr(\cdot)$ is a trace operator. Note that fidelity is bounded from $0$ to $1$, i.e., $0 \leq \Omega(\psi_m, \psi_{m'}) \leq 1$ and $\Omega (\psi_m,\psi_{m}) =1$. Briefly, fidelity between the two filters goes to $1$ when the similarity between two filters increases. On the other hand, the fidelity goes to $0$ when the similarity between the two filters decreases.

Motivated by the concept of fidelity, we aim to diversify inter-feature. Fig.~\ref{fig:inter} illustrates the semantic of inter-feature.
It is clear that adjusting fidelity $\Omega(\psi_l, \psi_{l'})$ affects the variation of inter-features, i.e., the inter-feature distance enlarges when the fidelity between two filters is high (and vice versa).
Thus, the reverse fidelity (RF) regularizer $L_{RF}$ is designed which minimizes the fidelity between all PQCs. The regularizer is defined as follows,

\begin{equation}
    L_{RF} = {\frac{1} {M(M-1)}} \sum_{m=1}^M \sum^M_{m'\neq m} \Omega(\psi_{m}, \psi_{m'}), \label{eq:L-RF}
\end{equation}
and note that the inter-feature becomes diverse, by minimizing $L_{RF}$. We will corroborate this in Sec.~\ref{sec:experiment}.

Algorithm~\ref{alg: RF-train} represents the overall training process of sQCNN-3D with RF-Train. The set of input point clouds and their labels are denoted as $(\mathbf{x},y)$. With the RF regularizer to achieve various features, we utilize cross-entropy loss $L_{CE}$ for classification, written as 
\begin{equation}
L_{CE} = -{\frac{1} {C}}\sum^C_{c=1} \log{p(y_{pred} = y_c|\mathbf{x})}, \label{eq:L-CE}
\end{equation}
where the number of classes is denoted as $C$. $y_{pred}$ and $y_c$ are predicted and actual classes. Finally, the train loss of sQCNN-3D with RF-Train is as,
\begin{equation}
    L_{train} ={\frac{1}{|\zeta|}} \sum_{(\mathbf{x}, y) \in \zeta} [L_{CE} + \lambda L_{RF}], \label{eq:L-train}
\end{equation}
where $\zeta$ and $\lambda$ denote a batch and an RF regularizer parameter.

\begin{algorithm2e}[t]
\small
    \SetCustomAlgoRuledWidth{0.44\textwidth}  
\caption{sQCNN-3D Quanvolution Procedure.}
\label{alg:local} 
\textbf{Notation.} Input point cloud $\mathcal{P}$, voxel $\mathcal{V}$ input feature $\mathcal{F}$ and filter size $\kappa$\;
\textbf{Input:} Input point cloud $P$\;
\textbf{Initialize.} $\mathcal{F} \leftarrow g(\mathcal{P}\odot\mathcal{V})$\;
    \For{$w \in \{1, 2, \cdots,{\hat{W}}\}$}
     {
         \For{$h \in \{1, 2, \cdots, {\hat{H}} \}$}
            {    
                \For{$ d \in \{1, 2, \cdots, {\hat{D}}\}$}
                {
                    Initialize quantum state, $\ket{\psi} \leftarrow \ket{0}$\;
                \For{$c \in \{1,2, \cdots,  c_{\text{in}}\}$}
                    {
                    Prepare data, $\mathbf{F}_{w:w+\kappa, h:h+\kappa, d:d+\kappa}$\;
                    
                    Upload data, $\ket{\psi} \leftarrow   U(\mathbf{F}_{w:w+\kappa, h:h+\kappa, d:d+\kappa}) \cdot \ket{0}^{\otimes q}$\;
                    Quanvolve data, $\ket{\psi} \leftarrow U(\bm \theta)\cdot \ket{\psi}$\;}
                    \For{$c \in \{1,2, \cdots,  c_{\text{out}}\}$}
                    {
                    Measure the output, $\langle O \rangle =\langle \psi| \mathbf{I}^{\otimes n-1}\otimes \mathbf{Z} |\psi\rangle$\;
                    }
                }
             }
 }
\textbf{Output:} Extracted features
\end{algorithm2e}

\subsection{sQCNN-3D Overall Algorithmic Procedure}\label{sec:training}
The Algorithm~\ref{alg:local} and Algorithm~\ref{alg: RF-train} represent the quanvolution process and training process of sQCNN-3D, respectively. In general, it is obvious that point cloud classification tasks require high computational and memory resources because of the massive data and input size of the point clouds. In addition, many qubits are needed to encode the entire point clouds into qubits, which can induce a barren plateau, in order to perform point-wise point cloud classification. As the solution under quantum computing concepts, the overall algorithmic procedure of sQCNN-3D are formalized as follows.
\begin{enumerate}
    \item 
    Embeds input point cloud $\mathcal{P}$ to geometric feature set $\mathcal{F}$.
    \item 
    Encodes the geometric feature set to patch with origin $(w, h, d)$ and kernel size $\kappa$. After computing the encoder, sQCNN-3D achieves $\mathbf{F}_{w:w+\kappa, h:h+\kappa, d:d+\kappa}$ in each stride.
    \item 
    Reuploads the input $\mathbf{F}_{w:w+\kappa, h:h+\kappa, d:d+\kappa}$ of each stride to quanvolutional filter.
    \item 
    Quanvolves the data through the quanvolutional filter.
    \item 
    Measuring and pooling occurs after quanvolution.
    \item 
    Concatenates the measured observable, i.e., extracted features, of each filter and exploit the data using FCN.
\end{enumerate}

\begin{algorithm2e}[t]\label{alg: RF-train}
\small
    \SetCustomAlgoRuledWidth{0.44\textwidth}  
\caption{sQCNN-3D with RF-Train.}\label{alg:Training algorithm} 
\textbf{Initialization.} sQCNN-3D parameters \;
 \For{ $e = \{1,2,\dots, E\}$}
 {
     \For{ $(\mathbf{x}, y) \in \zeta$}
     {  
        \For{ $j \in {[1, 2, \dots, \hat{W}\times\hat{H}\times \hat{D}]}$}
        {
            Get input feature $\mathbf{F}_{w:w+\kappa, h:h+\kappa, d:d+\kappa}  $\; 
            \For{$l, l^{'} \in \{1,2,\dots,L-1\}$}
             {
                Proceed quanvolution in Algorithm~\ref{alg:local}\;
             } 
         Calculate the output logits and cross-entropy loss with \eqref{eq:L-CE}\;
         Calculate RF-train regularization with \eqref{eq:L-RF}\;
         Obtain the train loss and its loss gradient\;
         Update trainable parameters of sQCNN-3D\;
        }
     }   
}
\end{algorithm2e}

\section{Performance Evaluation}\label{sec:experiment}
\subsection{Experimental Setting}

\begin{table}[t!]
    \caption{List of simulation parameters.}
\small
    \centering
    \resizebox{\columnwidth}{!}{
    \begin{tabular}{l|r}
    \toprule[1pt]
      \bf{Description}                & \bf{Value}  \\ \midrule
        \# of filters       & \{1, \textbf{2}, 4, 6, 8\} \\
        Size of voxels ($|\mathcal{V}|$)                & $32^3$ \\
        Optimizer                  & Adam \\
        Initial learning rate      & $0.001$ \\
        \# of qubits in a quanvolutional filter              & 4\\
        \# of params in a {scalable quanvolutional filter}    & $48$\\
        \# of params in a {classical filter}    & $64$\\
        Kernel size    & {$2^3$, $3^3$, $4^3$}\\
        Batch size                    & $8$\\
        Test batch size               & $128$\\
        \bottomrule[1pt]
    \end{tabular}
    }
    \label{tab:tab_parameters}
\end{table}
To corroborate the performance of the sQCNN-3D with RF-Train, the experiments are designed as follows.
\begin{itemize}
    \item First of all, we benchmark sQCNN-3D whether it achieves convincing accuracy as much as CNN in point cloud data classification. Our proposed sQCNN-3D with fine-tuned RF-Train is compared with classical CNN and QCNN algorithms.
    \item Second, we conduct an intensive experiment whether the diversity of extracted features can be increased by adjusting RF regularizer parameter $\lambda$, and it improves the performance. For this, we calculate the Euclidean distance between inter-features.
    \item Third, we compare the feature maps of sQCNN-3D and vanilla QCNN. Based on the visualization results as shown in Tab.~\ref{fig:Featuremap}, we corroborate that our proposed sQCNN-3D with RF-Train can fully exploit the intrinsic feature of each datum.
    \item Lastly, we observe the scaling strategy of sQCNN-3D. 
    We compare sQCNN-3D with various $\kappa$ to sQCNN-3D with fixed $\kappa$. We investigate the top-1 accuracy of two strategies. 
\end{itemize}

For the numerical experiments, we set the number of filters as 2 in both sQCNN and CNN models. In addition, we adopt top-1 accuracy for the metric. In addition, all quanvolutional filters are operated in 4 qubits system, i.e., $q=4$. The simulation parameters are listed in Tab.~\ref{tab:tab_parameters}. 
Note that we use the notation `{\footnotesize \texttt{Dataset (\# of classes)}}' to represent the type of dataset and the number of classes in experimental results.

 \begin{figure}[t]
    \centering
    \begin{tabular}{@{}c@{}c@{}}
         \includegraphics[width=.4999\columnwidth]{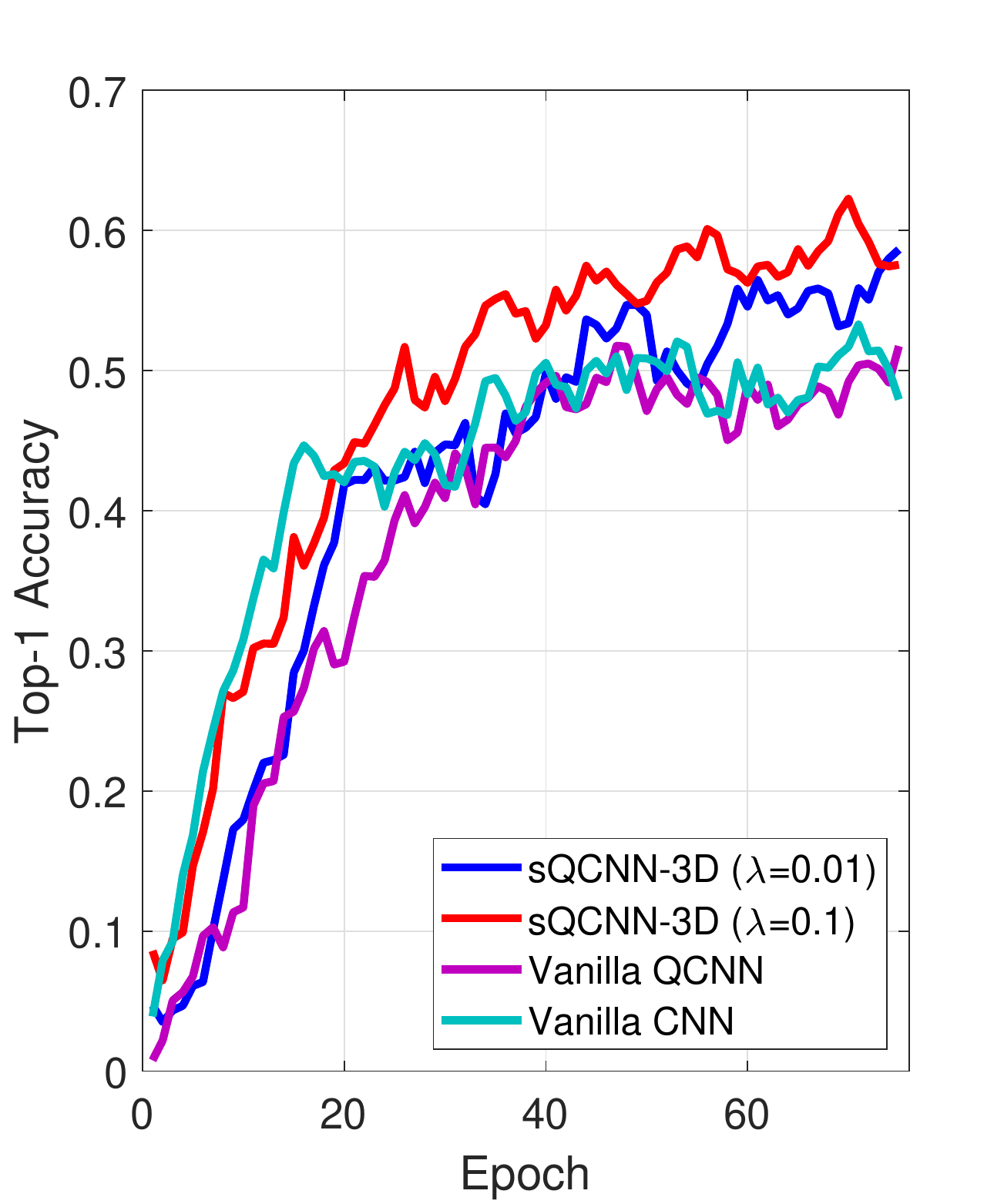}&
         \includegraphics[width=.4999\columnwidth]{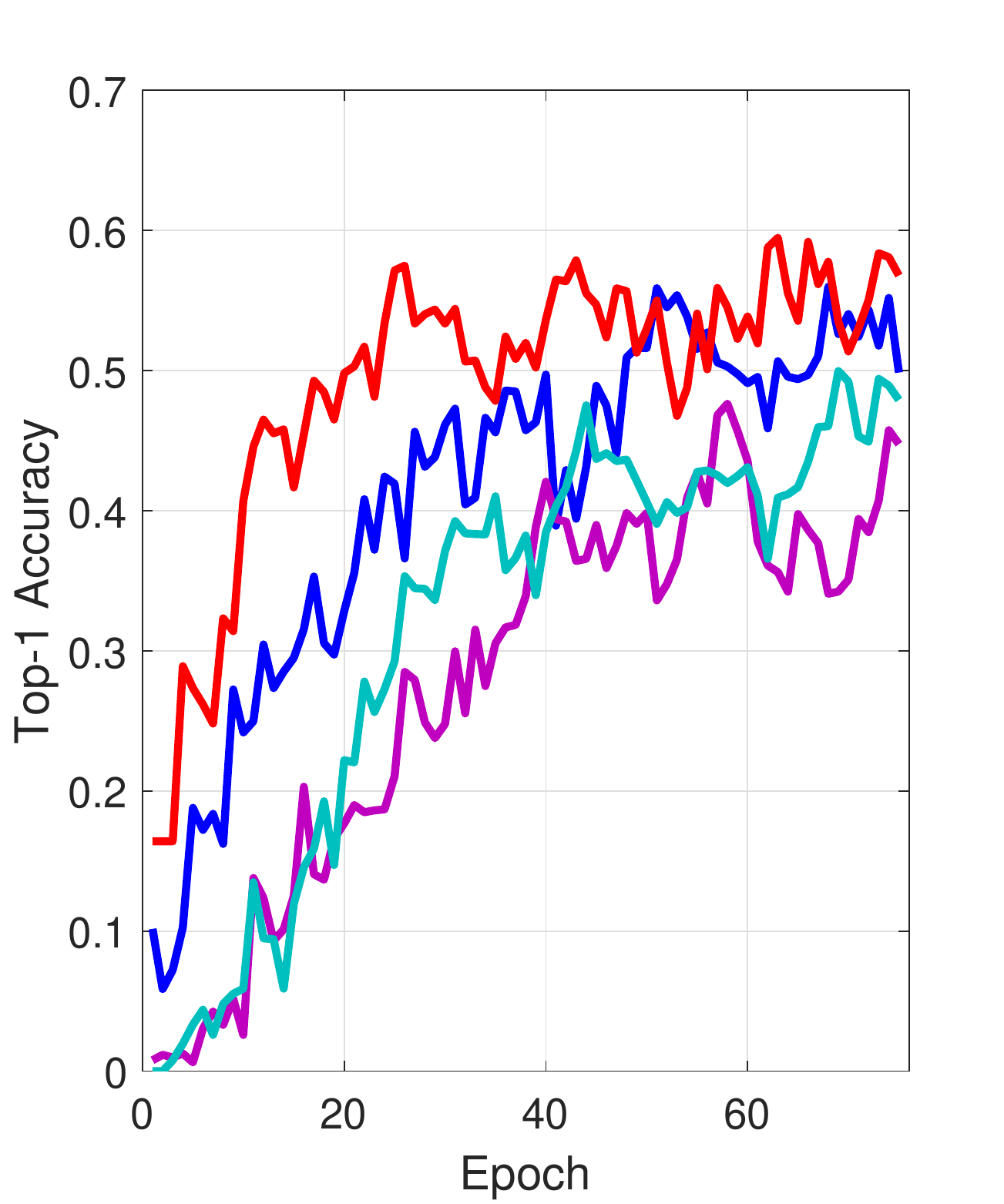}\\
         \small (a) ModelNet (40). & \small (b) ShapeNet (55).\\
    \end{tabular}
    \vspace{-3.5mm}
    \caption{Top-1 accuracy of sQCNN-3D with two datasets.}
    \label{exp:main}
\end{figure}

\begin{table}[t!]
\centering
\caption{Top-1 accuracy and Euclidean distance with ShapeNet (55).}\label{exp:rf-train}
\resizebox{\columnwidth}{!}{
\begin{tabular}{c||ccc}
    \toprule[1pt]   
     &\multicolumn{3}{c}{\textbf{RF-Train}}\\
    \textbf{Metric} & $\lambda=0$ & $\lambda=0.01$   &  $\lambda=0.1$ \\\midrule
    Top-1 accuracy (\%) & $39.5$& $53.14$& $\mathbf{55.3}$ \\
    Inter Feature distance ($\times 10^{-3}$) & $0.58$& $0.96$& $\mathbf{2.25}$ \\
\bottomrule[1pt]
\end{tabular}}\label{tab:distance} 
\end{table}

\subsection{Experimental Results}
\BfPara{Performance of sQCNN-3D}
Fig.~\ref{exp:main}(a)/(b) represent the point cloud classification with ModelNet40 and ShapeNet~\cite{wu20153d}. In both Fig.~\ref{exp:main}(a)/(b), sQCNN-3D with RF-Train outperforms vanilla CNN and vanilla QCNN. Especially, sQCNN-3D with RF regularizer parameter $\lambda=0.1$ achieves about $5.8\%$ and $12.9\%$ higher top-1 accuracy than QCNN with ModelNet40 and ShapeNet, respectively. In addition, we observe that sQCNN-3D achieves a lot higher top-1 accuracy than the CNN with 2 filters. It is the result of entanglement in quantum information theory~\cite{bennett1998quantum}. In contrast to CNN that conducts the dot product between the independent constituents of the input data and filter, sQCNN and QCNN entangle the input data simultaneously and it makes sQCNN and QCNN have more spatial information than CNN.

\begin{table}[t!]
\caption{
Extracted inter-feature maps according to various train strategies (i.e., untrained sQCNN-3D, Vanilla-Trained sQCNN-3D, and RF-trained sQCNN-3D).}\label{fig:Featuremap}
\centering
\includegraphics[width=\columnwidth]{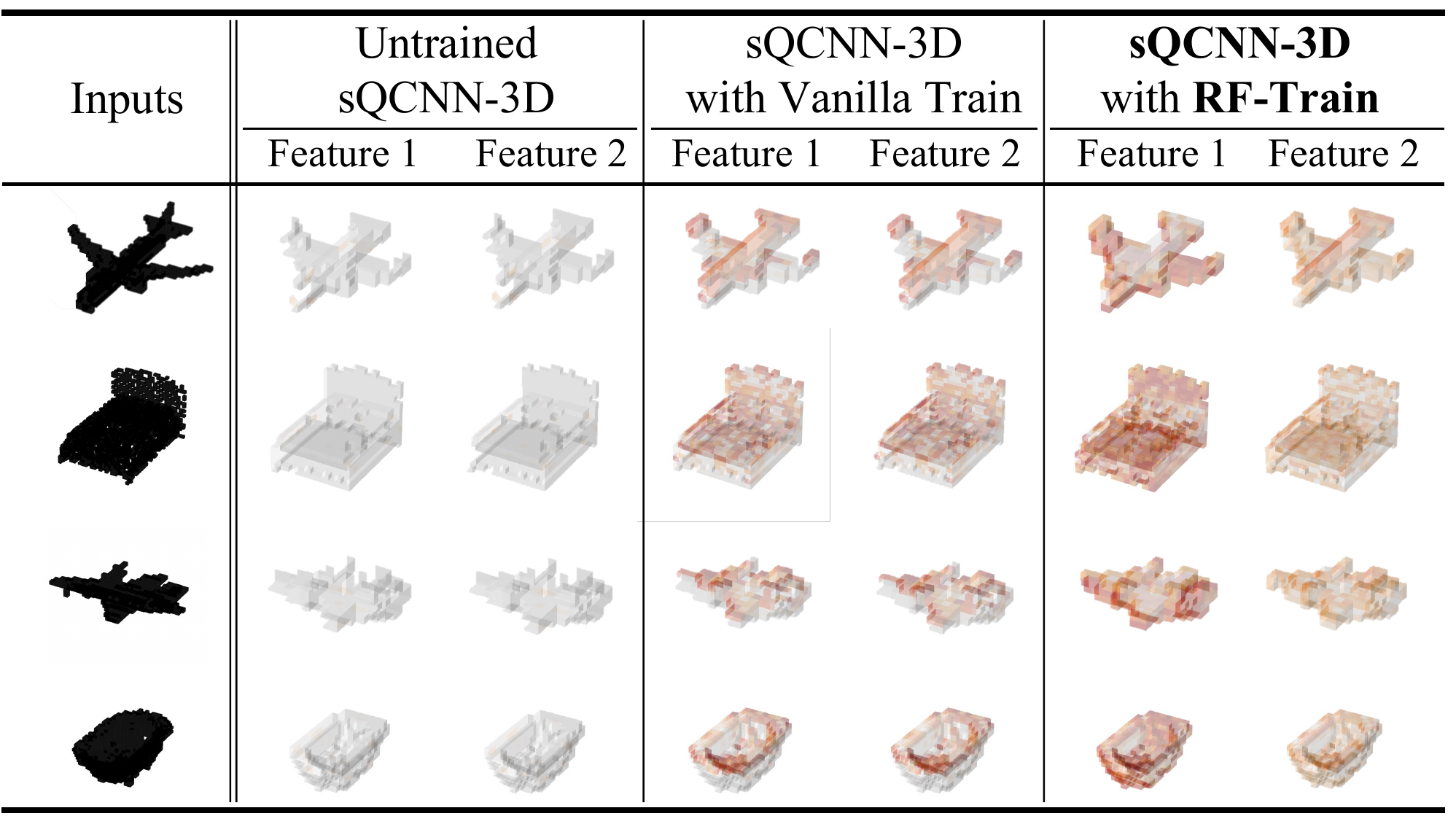}
\end{table}
\begin{figure}[ht]
    \centering
\includegraphics[width=\columnwidth]{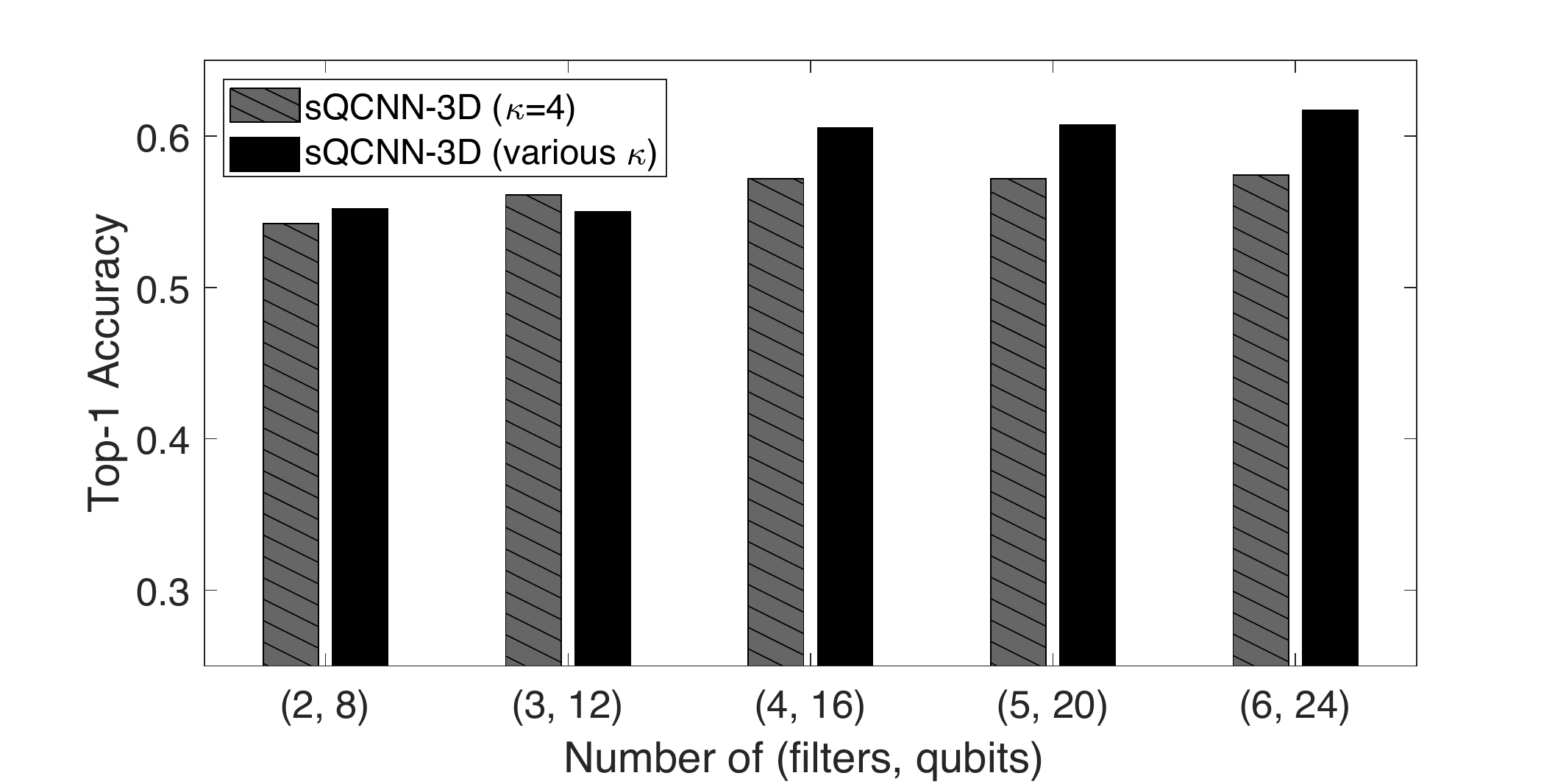}
    \caption{Scalability of sQCNN-3D with ShapeNet (55).}
    \label{exp:scalability}
\end{figure}

\BfPara{Effectiveness of RF-Train}
Tab.~\ref{exp:rf-train} shows the inter feature distance and top-1 accuracy according to various RF regularizer parameter $\lambda$. We observe that as the RF regularizer parameter $\lambda$ increases, inter feature distance increases, which leads to classification performance improvement. When the RF regularizer parameter $\lambda$ increases from $0$ to $0.1$, inter feature distance increases from $0.58$ to $2.25$ and the top-1 accuracy increases from $39.5\%$ to $55.3\%$. With a limited number of qubits (features), we observed that diversity between the extracted features is significant in the classification task, and RF-Train can increase this diversity. Further to this, we visualize the impact of Vanilla-Train and RF-Train on feature diversity. Tab.~\ref{fig:Featuremap} illustrates the individual feature maps extracted from untrained, vanilla-trained, and RF-trained filters. In contrast to the features extracted from vanilla-trained filter that show little diversity, the features extracted from RF-trained filters achieve reasonable diversity. Therefore, our proposed sQCNN-3D with RF-Train is able to exploit abundant intrinsic feature information, and this leads to performance improvements in classification tasks.

\BfPara{Scaling Strategy of sQCNN-3D} 
Fig.~\ref{exp:scalability} represents the scalability of CNN and sQCNN-3D with various random filter sizes ($\forall \kappa \in \mathbb{N}[2,4]$). Note that barren plateaus occur when the number of qubits increases in the quantum system. However, sQCNN-3D avoids the barren plateaus by increasing the number of filters while maintaining the number of qubits in a filter, even the number of (filters, qubits) increases from $(2, 8)$ to $(6, 24)$. In addition, we observe that combining several different kernel size $\kappa$ simultaneously when scaling up the number of  filters leads to performance improvements. The sQCNN-3D with various kernel sizes $\kappa$ achieves $6.5\%$ of top-1 accuracy improvement when the number of filters increases from 2 to 6, in contrast to sQCNN-3D with single kernel size $\kappa=4$ only achieves $3.2\%$ of top-1 accuracy improvement.  

\BfPara{Robust Performance on the Large Number of Classes} Tab.~\ref{exp:multiclass} shows the top-1 accuracy according to the number of classes. sQCNN-3D achieves $93.7\%$ top-1 accuracy on ModelNet with 10 classes. Especially, sQCNN-3D maintains the highest performance even if performance degradation occurs as the number of classes increases. In contrast to Vanilla QCNN and Vanilla CNN which show $46.8\%$ and $37.4\%$ of performance degradation, sQCNN-3D presents only $25.6\%$ of performance degradation on ShapNet. In addition, Vanilla QCNN and Vanilla CNN which show $36.14\%$ and $44.2\%$ of performance degradation, respectively, sQCNN-3D only reveals $34.3\%$ of performance degradation on ModelNet.

\section{Conclusions and Future Work}\label{sec:conclusions} 
\begin{table}[t!]
\centering
\caption{Top-1 accuracy according to number of classes.}\label{exp:multiclass}
\small
\begin{tabular}{c||ccc}
    \toprule[1pt]   
      Dataset& \multicolumn{3}{c}{\textbf{Top-1 Accuracy }}\\
     (\# of classes) &Vanilla CNN &Vanilla QCNN&sQCNN-3D\\\midrule[0.5pt]
    ModelNet (10)  &$90.3$& $89.73$ & $\mathbf{93.7}$ \\
    ModelNet (20)  &$74.2$& $83.28$ & $\mathbf{85.0}$  \\
    ModelNet (40)  &$46.1$& $53.59$ & $\mathbf{59.4}$ \\\midrule
    ShapeNet (10)  &$77.6$& $79.8$& $\mathbf{80.9}$  \\
    ShapeNet (30)  &$63.7$& $66.5$& $\mathbf{68.3}$ \\
    ShapeNet (55)  &$41.8$& $42.4$& $\mathbf{55.3}$  \\
\bottomrule[1pt]
\end{tabular}
\end{table}

This paper proposes a new 3D scalable quantum convolutional neural network (sQCNN-3D) for point cloud data classification. To fully exploit intrinsic features in point cloud data, which QCNN cannot, we designs a new sQCNN-3D in order to realize scalable 3D quanvolutional filtering functionalities. In addition, this paper proposes RF-Train and a novel scaling strategy to maximize the training performance of sQCNN-3D. With data-intensive performance evaluation, we corroborate that our proposed sQCNN-3D outperforms CNN and QCNN algorithms. We look forward to sQCNN-3D playing a leading role in point cloud data classification approaches in the near future based on its scalability and performance. Aligned with the development of quantum computing methodologies, it is obvious that the capabilities and scopes of sQCNN-3D will increase. Thus, it will be interesting to study the broad applications of sQCNN-3D extensions to large-scale point cloud data processing tasks such as object detection requiring high-accurate and high-speed data-intensive real-time computation.

\section*{Acknowledgment}
This research was funded by National Research Foundation of Korea (2022R1A2C2004869 and 2021R1A4A1030775). Won Joon Yun and Joongheon Kim are  the corresponding authors of this paper.

\bibliographystyle{IEEEtran}
\balance
\bibliography{PTCL}

\end{document}